\RequirePackage{fix-cm}
\documentclass[twocolumn]{svjour3}          
\smartqed  
\usepackage{graphicx,upgreek,color, url}
\usepackage{amsmath, amssymb}
\graphicspath{{Figs/}}
\newcommand{\gdot}{\ensuremath{\dot{\gamma}}} 
\newcommand{\PHIJ}{\ensuremath{\phi_{\rm J}^{\mu}}} 
\newcommand{\ZJ}{\ensuremath{Z_{\rm J}^{\mu}} } 

 \journalname{Granular Matter}
\begin{document}
\title{Force chains and networks: wet suspensions through dry granular eyes
}


\author{Rangarajan Radhakrishnan$^{1}$         \and
        John R. Royer $^{2}$\and
	Wilson C. K. Poon $^{2}$\and 
	Jin Sun$^{1}$ 
}


\institute{John R. Royer \at
	    \email{john.royer@ed.ac.uk}\\
 {}$^1$ Institute of Infrastructure and Environment, School of Engineering, The University of Edinburgh, Edinburgh EH9 3JL, UK. 
           \and
	   \at {}$^2$ SUPA and School of Physics and Astronomy, The University of Edinburgh, Edinburgh EH9 3FD, UK. \\
}

\date{Received: date / Accepted: date}

\maketitle

\begin{abstract}
	Recent advances in shear-thickening suspension rheology suggest a relation between (wet) suspension flow below jamming and (dry) granular physics. To probe this connection, we simulated the contact force networks in suspensions of non-Brownian spheres using the discrete element method (DEM), varying the particle friction coefficient and volume fraction. We find that force networks in these suspensions show quantitative similarities to those in jammed dry grains. As suspensions approach the jamming point, the extrapolated volume fraction and coordination number at jamming are similar to critical values obtained for isotropically compressed spheres. Similarly, the shape of the distribution of contact forces in flowing suspensions is remarkably similar to that found in granular packings, suggesting potential refinements for analytical mean field models for the rheology of shear thickening suspensions.
\end{abstract}
\section{Introduction}
Suspensions of non-Brownian particles (diameters $d\gtrsim 5{\upmu}$m) are ubiquitous in industry. Their rheology is critical in the manufacture of numerous formulated products, including paints, ceramics, and cosmetics. Provided the particles are well-stabilised, such suspensions typically shear thicken under flow, where the viscosity $\eta$ increases with increasing stress (or shear rate) \cite{Barnes1989,Denn2017}. There is a growing consensus \cite{Denn2017,Morris:2018aa,Guazzelli2018}, based on recent experiments and simulations \cite{Seto2013,Fernandez2013,Mari2014,Lin:2015aa,Clavaud:2017aa,Comtet2017}, that shear thickening results from the formation of frictional contacts. Friction constrains sliding motion, shifting the jamming volume fraction $\phi_{\rm J}$ from random close packing, $\phi_{\rm RCP}$, for frictionless particles at low stress to $\phi_{\rm m} < \phi_{\rm RCP}$ for frictional particles at high stress.  Full flow curves $\eta(\sigma)$ can be calculated using the phenomenological Wyart-Cates (WC) model \cite{Wyart2014}, in which the stress-dependent $\phi_{\rm J}(\sigma)$ interpolates between these two limits, reproducing both simulations and experimental results \cite{Guy2015,Royer2016,Singh2018a}.  

The importance of static friction and constraints in shear thickening \cite{Hecke:2010aa} suggests an intimate connection with the physics of dry granular media. 
Similar features in the spectrum of vibrational modes (density of states) in suspensions and jammed grains have been shown in simplified simulations of frictionless suspensions \cite{Lerner2012a,Lerner2012b,During2014}, with low frequency `soft modes' that vanish at a critical coordination number $Z \rightarrow Z_{\rm J}$, driving the viscosity divergence. This result and the reduction of $Z_{\rm J}$ in frictional packings \cite{Hecke:2010aa,Silbert2010} underpins the initial WC formulation \cite{Wyart2014}, with $\phi_{\rm m}$ in suspensions corresponding to the limit of random loose packed frictional grains. 
 
Motivated by a lack of obvious structural signatures of jamming in traditional real-space measures, there has been considerable effort devoted to characterizing the statistics and structure of the contact force networks of dry granular materials \cite{Radjai:1996aa,Mueth1998,Makse:2000aa,Blair2001,Snoeijer:2004aa,Majmudar:2005aa,Corwin:2005kk}. With limited exceptions \cite{Melrose:2004ab}, contact networks and force distributions in suspensions have received less attention. However, in a `granular view' of shear thickening, force distributions are key \cite{Seto2013,Mari2014}, particularly at the thickening transition where the form of the force distribution governs the fraction of frictional contacts at a given stress \cite{Guy2015,Ness2016b}.

Here we probe this connection between suspensions and dry grains near jamming in detail, using DEM simulations to model suspensions of spherical particles at varying volume fractions $\phi$ and inter-particle friction coefficients $\mu$. Computing the mean coordination number and force distributions in these simulations, we uncover numerous similarities between flowing suspensions near jamming and dry granular packs, supporting the `granular view' of suspension rheology.

\section{Methodology}
We simulated 3D suspensions of neutrally-buoyant non-Brownian spheres in a Newtonian background fluid of viscosity $\eta_{\rm f}$. Hydrodynamic forces are calculated using the discrete element method (DEM)~\cite{Ball1997,Seto2013,Ness2015}, including the Stokes drag and short-ranged lubrication interactions. The latter are regularized below a particle separation of $\xi_{\rm min} $ where $\xi = 2 r/(d_1+d_2)$ for two particles of diameter $d_1$, $d_2$ at a center-to-center distance $r$~\cite{Cheal2018,Radhakrishnan2018}. We neglect long-ranged hydrodynamic interactions between particles because we only consider high $\phi$. 

The regularization of lubrication forces means that particles can come into contact. Such contacts are modeled by a linear Hookean spring $F_{\rm n}^c=k_{\rm n} h$, with spring constant $k_{\rm n}$  and the extent of particle overlap $h$. The tangential spring force is $F_t^c=k_t h_t$, where $k_t=(2/7)k_{\rm n}$ and the incremental tangential stretch $h_t$ is initialized to 0 at contact and updated following Silbert et al.~\cite{Silbert2001}. The maximum $F_t^c$ is set to satisfy the Coulomb criterion, $|F^c_t| \le \mu |F_{\rm n}^c|\ ,$ where $\mu$ is the friction coefficient. 

The critical load model (CLM), in which $F_t^c=0$ when $F_{\rm n}^c \le F_{\rm CLM}$, is used to simulate shear thickening~\cite{Seto2013,Mari2014}.  $F_{\rm CLM}$ mimics the repulsion that prevents facile particle contact~\cite{Comtet2017,Chatte2018}, and sets a force scale for transition between frictionless to frictional flow.

Homogeneous simple shear at rate $\gdot$ was imposed by affine deformation with Lees-Edwards periodic boundary condition in LAMMPS~\cite{Plimpton1995}. 1872 bi-disperse spheres at diameter ratio 1:1.4 (to prevent crystallization) mixed in equal volumes were simulated~\cite{Seto2013} at constant $\phi$ and $\gdot$ such that the Stokes' number $\rho \dot{\gamma} d^2/\eta_{\rm f} < 1$, where $\rho$ is the fluid or particle density.  

The total stress was found by summing contributions from the contact, hydrodynamic forces between particles and isolated particle stresslets. The relative viscosity is taken to be $\eta_{\rm s}=\sigma/(\gdot \eta_{\rm f})$, where $\sigma$ is the $xy$ component of the stress tensor. All the results presented were obtained by averaging over at least 10 strain units in steady state. To ensure hard-sphere like behavior, in all simulations we verified that the stiffness was set sufficiently high, $k_{\rm n} \gg \sigma d$, to avoid spurious shear thinning from particle overlaps~\cite{Radhakrishnan2019}.

\section{Results}

\begin{figure}
\center
  \includegraphics[width=0.5\textwidth]{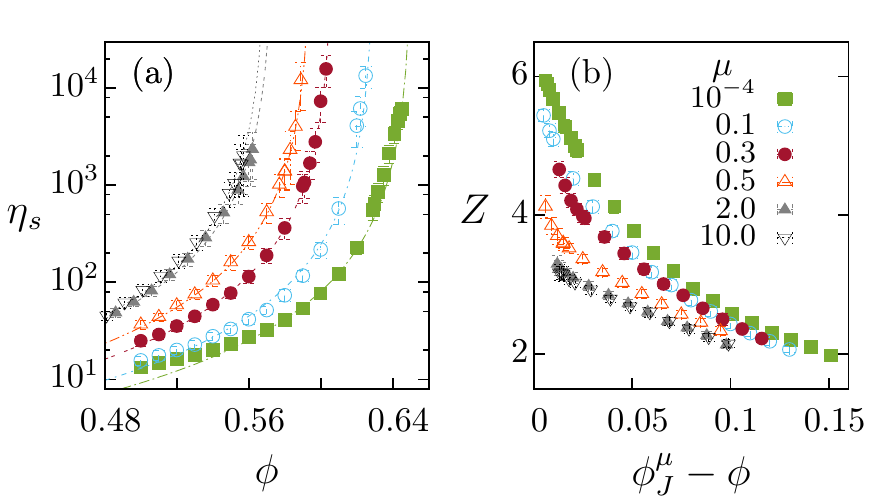}
\caption{(a) Relative viscosity $\eta_{\rm s}$ of suspensions at different volume fractions $\phi$ and (b) the average contact number $Z$ for different particle friction coefficients $\mu$. The jamming packing fraction \PHIJ is obtained by fitting $\eta_s$ with Eq.~\ref{eq:rel_vis} indicated by the dot-dashed lines in (a). }
\label{fig:rel_vis}      
\end{figure}

At fixed $\mu$, the suspension's shear rheology is quasi Newtonian. Its $\gdot$-independent viscosity increases with $\phi$ and diverges at a $\mu$-dependent jamming volume fraction \PHIJ, Fig.~\ref{fig:rel_vis} (a), which we obtain by fitting
\begin{equation}
	\eta_{\rm s}= \alpha^\mu \left(\PHIJ -\phi\right)^{p},\ 
	\label{eq:rel_vis}
\end{equation}
where, following literature \cite{Singh2018a}, we fix $p=-2$, leaving $\alpha^\mu$ and \PHIJ\, as fitting parameters.
As $\mu$ varies from $10^{-4}$ to 10, $\PHIJ$ moved from 0.65 in the low friction limit to 0.57 in the high friction limit, agreeing with previous work \cite{Mari2014,Singh2018a}. The pre-factor varies weakly with $\mu$ and, despite previous suggestion of $p$ varying somewhat from $-2$ in the frictionless limit~\cite{Kawasaki2015}, our results are consistent with $p=-2$ at all $\mu$ (Fig. S2~\cite{SI}).  Fits yielded nearly identical \PHIJ\, even if the $p$ is allowed to vary.

As $\phi \to \phi_{\rm J}^{\mu}$, the average per-particle contact (or coordination) number $Z$ increases, Fig.~\ref{fig:rel_vis} (b). We estimate $Z$ at jamming, \ZJ, by linearly extrapolating data for $\PHIJ-\phi\le0.011$ towards zero. Large fluctuations occurred for  $\mu>0.75$ and $\PHIJ-\phi \lesssim 0.01$ due to jamming and unjamming transitions in a finite-size system~\cite{Banigan2013}. Thus, we do not extrapolate to find \ZJ above  $\mu=0.75$. 

\begin{figure}
\center
  \includegraphics[width=0.5\textwidth]{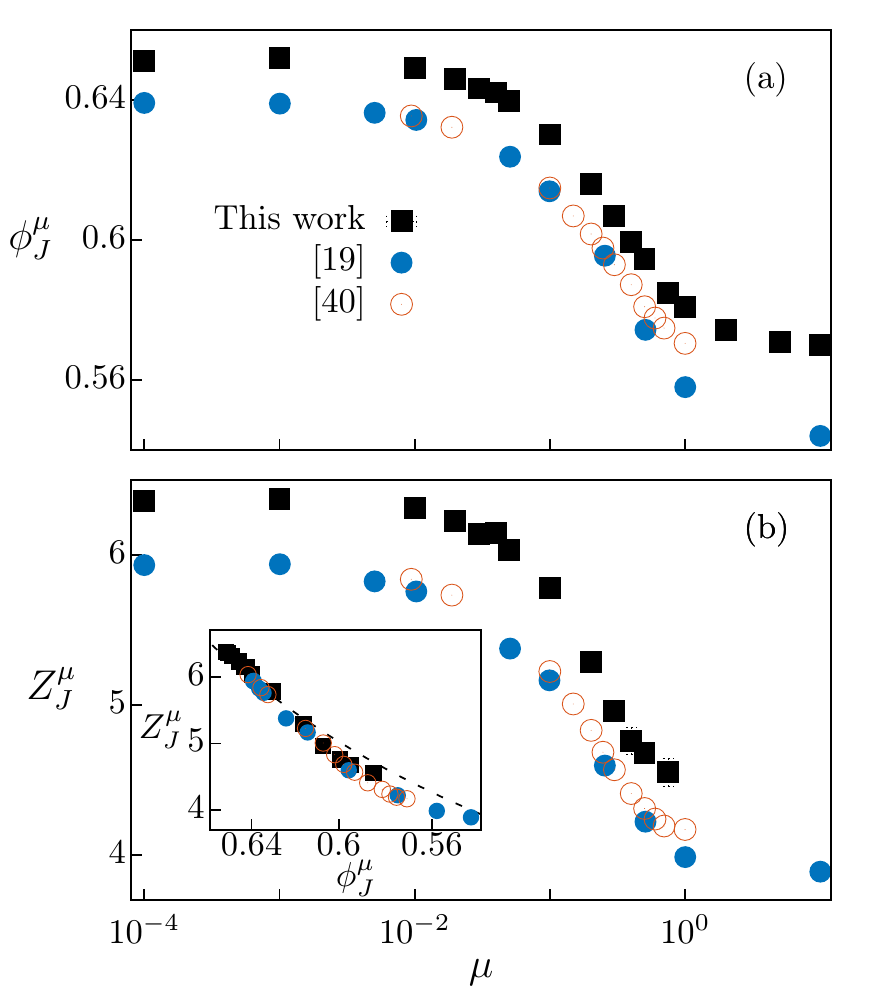}
  \caption{Dependence of (a) the critical jamming volume fraction \PHIJ, and (b) the critical contact number \ZJ of bidisperse suspensions with friction coefficient $\mu$ compared with granular monodisperse spheres~\cite{Silbert2010,Sun:11a}. The inset shows the relation between \ZJ and \PHIJ. The dashed line shows Eq. \ref{eq:phij_rel} with $q=1.676$. }
\label{fig:critical}       
\end{figure}

The plots of \PHIJ\, and \ZJ against $\mu$, Fig.~\ref{fig:critical}, show a remarkably similarity in form compared to simulations of isotropically-compressed packings \cite{Silbert2010} and simple shear \cite{Sun:11a} of monodisperse granular spheres.  Our \PHIJ\, values are slightly higher than those in monodisperse systems, as expected~\cite{Pednekar2018c}. The same is true of \ZJ even in the low-$\mu$ limit, where $Z^{\mu=0}_{\textrm{iso}} = 6$ is expected. It is not clear if the finite particle softness in DEM simulations or shear-induced anisotropic structures is responsible for this difference, which has also been observed for quasi-statically sheared bidisperse granular spheres~\cite{Vinutha2019}.

\begin{figure}
	\includegraphics[width=0.5\textwidth]{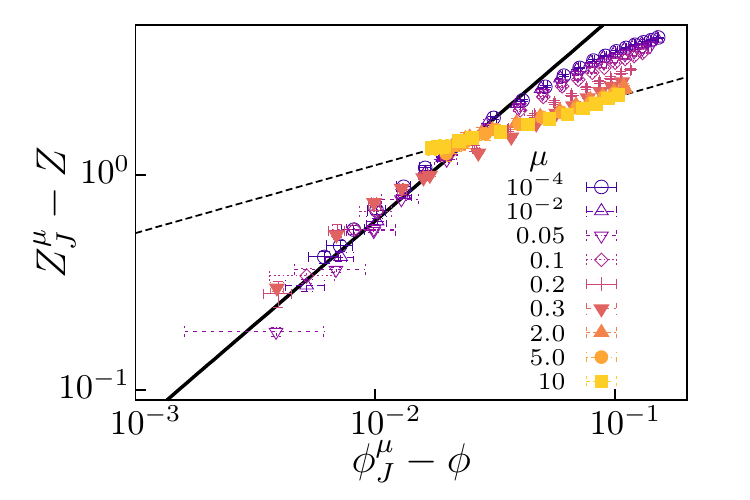}
	\caption{Contact number deficit $\Delta  Z = \ZJ-Z$ verses distance to jamming $\Delta \phi = \PHIJ-\phi$ for suspensions with varying friction coefficients $\mu$.  Lines: fits to $\ZJ-Z \sim (\PHIJ -\phi)^n$ giving $n=0.96$ (solid) and $n=0.32$ (dashed).}
\label{fig:Zpow}
\end{figure}

For monodisperse spheres, the Edwards statistical ensemble packing theory \cite{Edwards:1989} leads to \cite{Song2008,Wang2010,Baule2018} 
\begin{equation}
	\phi_{\textrm{RLP}}=\frac{Z}{Z+2q} 
	\label{eq:phij_rel}
\end{equation}
 with $q=\sqrt{3}\simeq 1.732$. Eq.~\ref{eq:phij_rel} fits our data with $Z \to \ZJ$, $\phi \to \PHIJ$ and $q=1.676$, Fig.~\ref{fig:critical} (inset). 
We plot the contact number deficit $\Delta Z \equiv \ZJ-Z$ against the distance to jamming $\Delta\phi\equiv\PHIJ-\phi$ in Fig.~\ref{fig:Zpow}. In compressed granular packings, one finds $Z-\ZJ\propto (\phi-\PHIJ)^{1/2}$ and results for $0 \leq \mu \leq 10$ overlap in the range $10^{-5} \lesssim \Delta \phi \lesssim 10^{-1}$. \cite{Silbert2010}. The situation in suspensions is more complex, Fig.~\ref{fig:Zpow}. For $\mu \lesssim 0.06$, our results collapse onto two power law regimes, following $\Delta Z\propto \Delta\phi^n$, with $n\approx 1$ below $\Delta\phi \approx 0.06$, and $n\approx 0.3$ beyond.

For $\mu \gtrsim 0.1$, the low $\Delta \phi$ data for different $\mu$ still collapse onto a power-law behavior with $n \approx 1$. Now, however, we no longer find collapse at high $\Delta\phi$. Instead, data at increasing $\mu$ deviate from the $n=1$ power law sooner and drop to a lower asymptote. Data at $\mu \gtrsim 2$ is subject to significant uncertainty because of the impossibility of extrapolating to find \ZJ in Fig.~\ref{fig:rel_vis}(b). Instead, we solve for \ZJ using Eq.~\ref{eq:phij_rel} by substituting values of $\PHIJ$\, obtained previously from fitting Fig.~\ref{fig:rel_vis}(a). Including the points so obtained, we find that the data sets for $\mu \gtrsim 0.1$ appear to converge toward a $n = 0.3$ power law, the same exponent as the low-$\mu$ data sets, but now with a lower amplitude. It is yet unknown if the power law differs at lower $\Delta \phi$ at the highest $\mu$.

\begin{figure*}
\centering
  \includegraphics{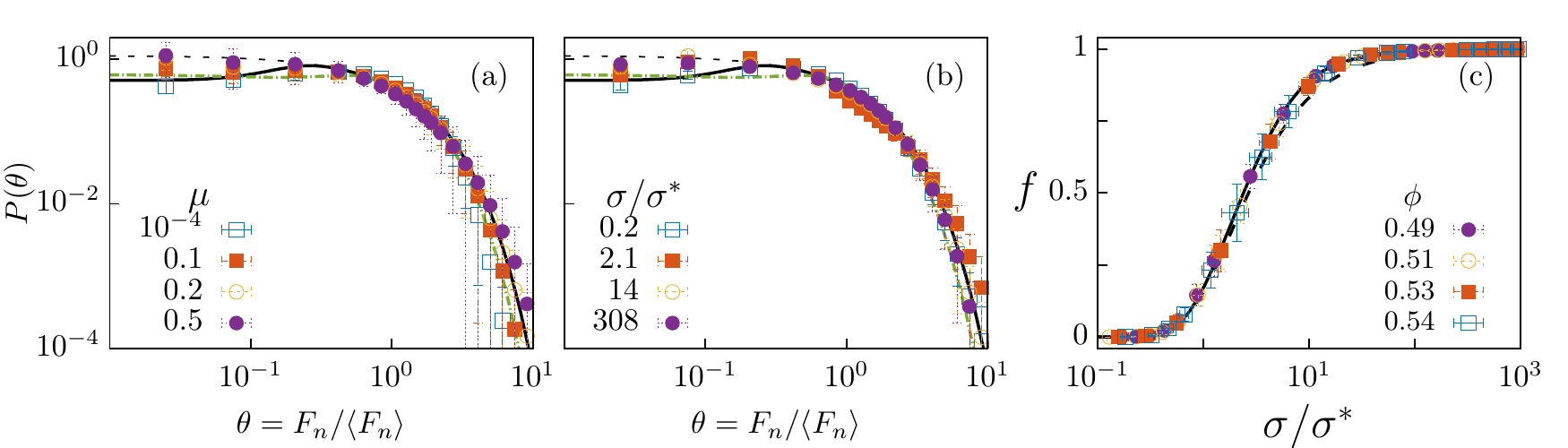}
 \caption{Probability density functions $P(\theta)$ of the total normal force at contact between particles of suspensions for: (a) different friction coefficients when $\PHIJ-\phi \approx5\times10^{-3}$, (b) different stresses under the critical load model (CLM) with fixed $\phi=0.54,\mu=0.5$. (c) The fraction of frictional contacts $f$ with increase in stress for suspensions at different $\phi$ with $\mu=0.5$ for the CLM. In  (a) and (b), fits to Eq.~\ref{eq:dist} for $\mu\to0$ (solid line),  $\mu\to\infty$ (dashed line), and those from previous experiments~\cite{Blair2001} (dot-dashed line) with parameters given in Table~\ref{tab:param} are shown. The solid and dashed lines in (c) are the $f$ obtained from Eq.~\ref{eq:deriv_dist} for the solid and dashed lines in (a) and (b) with $\alpha=1.85$.  }
 \label{fig:normal_dist}       
\end{figure*}

The rigidity of granular packs is mediated through grain-grain contacts, which are distributed heterogeneously in space \cite{Jaeger:1996fk}, with some grains bearing many times the mean force while others bear almost none. This is captured by the probability distribution $P(\theta)$, where $\theta \equiv F_{\rm n}/\langle F_{\rm n} \rangle$ is the normal force between neighbors $F_{\rm n}$ normalized by the mean. In dry granular packing, normal forces arise exclusively from direct contacts. In suspensions, inter-particle hydrodynamic interactions also contribute, although their exclusion does not change our results (Fig. S6~\cite{SI}).  We calculate $P(\theta)$ from snapshots of the force network collected every 0.1 strain units averaged over 10 strain units, with error bars indicating the standard deviation.

In Fig.~\ref{fig:normal_dist}(a) we plot $P(\theta)$ for sheared suspensions close to jamming with $\Delta \phi \simeq 5\times10^{-3}$ for a range of $\mu$. In common with dry granular packings, we find exponential-like tails at high forces. At $\mu =10^{-4}$, $P(\theta)$ is peaked at $\theta\approx 0.5$. As $\mu$ increases, the probability of low-force contacts increases, until at $\mu = 0.5$ the peak in $P(\theta)$ is no longer evident. An increase in $P(\theta \to 0)$ with increasing $\mu$ was also reported in simulations of dry granular packings~\cite{Silbert2010}. We observe a similar trend for fixed $\mu=0.2$ at varying $\phi$, with a slight increase in $P(\theta \to 0)$ as $\Delta \phi$ decreases (Fig. S3~\cite{SI}). Distributions of the tangential contact forces and the fraction of mobilized contacts are likewise similar to those obtained from dry-grain simulations (Fig. S4 ~\cite{SI}).

In compressed dry granular packings, $P( \theta)$ can be described by the empirical relation \cite{Mueth1998,Blair2001}
\begin{equation}
P( \theta)=a\left( 1 - b e^{-c  \theta^2} \right) e^{-\beta \theta}\ .
\label{eq:dist}
\end{equation}
We fit our data to this form, Table~\ref{tab:param}. The data and fit, Fig.~\ref{fig:normal_dist}(a) (full line), at $\mu =10^{-4}$ are more pronouncedly peaked than the $P(\theta)$ for compressed amorphous packings of smooth spheres, Fig.~\ref{fig:normal_dist}(a) (dot-dashed). At $\mu=0.5$, the peak in $P(\theta)$ is no longer evident: $b=0$ and the distribution is purely exponential, Fig.~\ref{fig:normal_dist}(a) (dashed). Note that our results at high $\mu$ show, and our results at low $\mu$ are consistent with, a finite plateau rather than a power-law scaling as $\theta\rightarrow 0$~\cite{Lerner2012b,Radjai:1996aa,Wang2010,Kyeyune2018}.

To examine the force network during shear thickening, we implement a critical load model (CLM), where frictional contacts are activated whenever $F_{\rm n}$ exceeds a threshold $F_{\rm CLM}$. 
For a shear-thickening suspension with $\phi=0.54,\mu=0.5$, our data for $P(\theta)$, Fig.~\ref{fig:normal_dist}(b), are almost identical to those at fixed $\mu$, Fig.~\ref{fig:normal_dist}(a), and can be fitted to Eq.~\ref{eq:dist} in the low- and high-stress limits using the previous low- and high-$\mu$ limit parameters (solid and dashed lines respectively).
In the CLM, $F_{\rm CLM}$ determines the onset stress $\sigma^* = F_{\rm CLM}/(1.5 \pi d^2)$ for shear thickening \cite{Seto2013,Mari2014,Ness2016b}. The WC jamming volume fraction $\phi_{\rm J}(\sigma)$ shifts from $\phi_{\rm J}^0$ to $\phi_{\rm J}^\mu$ as the fraction of frictional contacts $f(\sigma)$ increases from $f \to 0$ for $\sigma \ll \sigma^*$ to $f \to 1$ for $\sigma \gg \sigma^*$, giving a thickening flow curve $\eta(\sigma)$. 

\begin{table}
\caption{Fit parameters near jamming for $P(\theta)$}
\label{tab:param}       
\begin{tabular}{lllll}
\hline\noalign{\smallskip}
 & a & b & c & $\beta$  \\
\noalign{\smallskip}\hline\noalign{\smallskip}
$\mu = 10^{-4}$ & 1.12 & 0.54 & 33 & 1.0 \\
$\mu \to \infty$ & 1.12 & 0 & - & 1.0 \\
Amorphous smooth spheres~\cite{Blair2001} & 1.5 & 0.59 & 3.1 & 1.21 \\
\noalign{\smallskip}\hline
\end{tabular}
\end{table}
%


Typically, $f(\sigma) = \exp(-\tilde{\sigma}^*/\sigma)$ in fitting WC-type models to experiments or simulations~\cite{Guy2015,Singh2018a}, where $\tilde{\sigma}^* \approx \sigma^*$ to $\mathcal{O}(1)$. In general, the fraction of frictional contacts is given by
\begin{equation}
f=P(\theta \ge \theta_{\rm CLM})= \frac{\int_{\theta_{\rm CLM}}^{\infty} P(\theta)\, {\rm d}\theta}{\int_{0}^{\infty} P(\theta)\, {\rm d}\theta}\ ,
\label{eq:deriv_dist}
\end{equation}
where $\theta_{\rm CLM}(\sigma)=F_{\rm CLM}/\langle F_{\rm n}(\sigma) \rangle$. Assuming $\langle F_{\rm n} \rangle \propto \sigma$, we can write $\theta_{\rm CLM}=\alpha \sigma^*/\sigma$, where the pre-factor $\alpha \simeq 1.85$ is found to be independent of $\phi$ (see Fig. S7 and associated discussion~\cite{SI}). For a simple exponential force distribution, corresponding to our high-$\mu$ form of Eq.~\ref{eq:dist} with $b=0$ and $\beta=1$, integrating Eq.~\ref{eq:deriv_dist} gives $f(\sigma) = \exp(-\alpha \sigma^*/\sigma)$, a result previously derived to motivate an exponential form for $f(\sigma)$ ~\cite{Guy2015,Ness2016b}. This procedure can be repeated using the low-$\mu$ fitting parameters in Eq.~\ref{eq:dist}, where $b,c \neq 0$. Interestingly, the resulting $f(\sigma)$ obtained using either low- and high-$\mu$ fitted parameters in Eq. \ref{eq:deriv_dist}, corresponding to the solid  and dashed lines in Fig.~\ref{fig:normal_dist}(c), differ little; each gives a reasonable account of the data. This reflects the dominant importance of the high-force exponential tail, which remains largely unchanged as either $\mu$ or $\phi$ is varied.

\section{Discussion and  Conclusions}
We find that sheared dense suspensions exhibit a number of quantitative similarities to dry granular materials near jamming. The $\mu$-dependence of the volume fraction and coordination number at jamming, $\phi_{\rm J}^\mu$ and $Z_{\rm J}^\mu$, are reminiscent of corresponding functions in isotropic sphere packings~\cite{Silbert2010}. Thus, theoretical models relating $\phi_{\rm J}^\mu$ and $Z_{\rm J}^\mu$ in the latter may be extendable to suspensions~\cite{Song2008,Wang2010}. The force distribution in suspensions also recalls that in granular packings. An exponential tail dominates in the $\mu \to 0$ and $\infty$ limits, justifying the use of a simple exponential form for the fraction of frictional contacts in WC-type models of shear thickening. 

Despite these similarities, we find that the relation between $\Delta Z$ and $\Delta \phi$ is more complex in suspensions than in dry grains, undercutting a number of assumptions underpinning the WC model for shear thickening. As initially formulated \cite{Wyart2014}, WC drew upon simulations of frictionless hard spheres \cite{Lerner2012a,Lerner2012b}, which found $\eta \sim \Delta Z^{-\nu}$ with $\nu \sim 2$. The basic physics is that `soft modes') are lost as the system approaches isostaticity ($\Delta Z \to 0$). These soft modes are characterized by the vibrational density of states $D(\omega)$, which gives the number of modes per particle at a given frequency. In dry granular packings above jamming, $D(\omega)$ trasitions from classical Debye scaling ($\propto \omega^2$ in three dimensions) to a constant low-frequency plateau as $|Z - Z^{\mu}_{\rm J}| \to 0$ in both the frictionless and frictional case \cite{Hecke:2010aa,Henkes:2010aa}. A similar change in the shape of $D(\omega)$ was seen in simulated frictionless suspensions as $Z \to Z^{\mu=0}_{\rm J}$ \cite{Lerner2012a}.

WC assume that the same physics applies in wet frictional suspensions, so that the shape of $D(\omega)$ is likewise controlled by $Z^{\mu}_{\rm J}$ and the $\eta \sim \Delta Z^{-\nu}$ scaling applies in both the frictionless and frictional case. To make useful predictions, they then need to relate the `natural variable' in dry grains, $Z$, to the `natural variable' in suspensions, $\phi$, for which they make the further assumption that $\Delta Z \propto \Delta \phi$ in {\it both} the low- and high-$\mu$ limits. We find such proportionality only for $\mu \lesssim 0.06$ and very close to jamming ($\Delta\phi \lesssim 0.03$), Fig.~\ref{fig:critical}. In these low-friction suspensions there is a rollover to a weaker power law at higher $\Delta \phi$, and for higher values of $\mu$ we never reach a regime where $\Delta Z \propto \Delta \phi$. 

Despite these discrepancies, Eq. \ref{eq:rel_vis} fits our results over a relatively wide range of $\Delta\phi$ for all values of $\mu$, and the WC model formulated in terms of $\phi$ has proved successful in capturing experimental results \cite{Guy2015,Royer2016}. This suggests that the empirical relation between $\eta$ and $\Delta \phi$ should in fact be viewed as more universal, while the `soft mode' approach to understanding the viscosity divergence in suspensions may have limited applicability. Indeed, this is in line with work to develop empirical constitutive relations for frictional suspensions by analogy to empirical constitutive relations for flowing grains \cite{Boyer2011}. It has also recently been shown that viscosity divergence $\eta \propto \Delta\phi^{-2}$ can be derived for suspensions of frictionless spheres through a non-equilibrium kinetic theory approach~\cite{Suzuki2019}, and this approach could perhaps be extended to frictional suspensions as well.

In our analysis of the force networks in these suspensions, we have neglected the contribution from contact anisotropy to the stress and possible microstructure evolution during shear thickening. However, given the similarities between the force networks in wet suspensions and dry grains, it is likely that more advanced methods used to characterize force networks in jammed packings \cite{Sarkar:2013aa,Thomas2018,Kollmer:2019aa} could be applied to open a new window into the rheology and dynamics of dense suspensions.

\begin{acknowledgements}
The UK EPSRC Future Formulations grant EP/N025318/1 funded this work.
\end{acknowledgements}


%
\end{document}